\documentstyle[aps,preprint,graphicx]{revtex}
\begin{document}


\noindent
{\bf
Momentum creation by vortices in $^3$He experiments
as a model of primordial baryogenesis.}\\

\noindent
 T.D.C. Bevan$^1$, A.J. Manninen$^{1,2}$, J.B. Cook$^1$,
J.R. Hook$^1$, H.E. Hall$^1$, T. Vachaspati$^3$ and G.E.
Volovik$^{4,5}$ \\
\footnotetext{
$^1$ Schuster Laboratory, University of Manchester,
Manchester, M13 9PL, UK\\
$^2$ Now at Physics Dept., University of Jyv\"askyl\"a,
P.O.Box35, 40351
Jyv\"askyl\"a, Finland\\
$^3$ Physics Department, Case Western Reserve University,
Cleveland, OH 44106, USA\\
$^4$ Low Temperature Laboratory, Helsinki University of
Technology, 02150 Espoo, Finland\\
$^5$ Landau Institute for Theoretical Physics, 117334
Moscow, Russia. }

\noindent



{\bf
An important problem in cosmology is why the universe contains so
much more matter than antimatter. Such an imbalance can result from
processes in which baryon number is not conserved. These may
have occured during the electroweak phase transition in the early
universe when elementary particles first acquired 
mass\cite{Dolgov,Turok,Review2}. 
The standard model of particle physics has been thoroughly
tested in terrestrial accelerator experiments where relatively
low energies are involved and perturbation theory can be used to
make predictions.  But these experiments and calculations are unable
to probe the processes expected to occur during the electroweak
phase transition, which involve large departures from the
vacuum state.  It is worthwhile therefore to
investigate systems where similar processes can be studied
experimentally. Superfluid $^3$He has many similarities to the
standard electroweak model\cite{VolovikVachaspati}. 
The analogue of cosmic string production 
has already been demonstrated in condensed 
matter\cite{Zurek,nematic,helium,BigBangNature1,BigBangNature2}. 
Here we explain how recent $^3$He experiments at Manchester
demonstrate the creation 
of excitation momentum (momentogenesis) by quantized vortices, 
a process analogous to baryogenesis within cosmic strings.}



To explain the creation of matter in the early universe
we begin by recalling the relationship
$E^2=m_0^2c^4+p^2c^2$
between energy $E$ and momentum $p$ for relativistic particles
with rest mass $m_0$. Dirac realised that the square root of this
equation $E=\pm\sqrt{m_0^2c^4+p^2c^2}$ produced particles with
both positive and negative energy. This led to his famous picture
of the vacuum state (the state with no real particles) as one in
which all the negative energy states are full and all the positive
energy states are empty. A real particle is then created as shown
in Fig. 1(a) by excitation of a particle from the ``Dirac sea" of
negative energy states to a positive energy state. Such processes
however create matter and antimatter in equal amounts since the
appearance of a hole in the negative energy states is interpreted
as the simultaneous creation of an antiparticle.

The net creation of matter in the form of baryons such as protons
and neutrons requires processes in which antiparticles are not
simultaneously created and thus
baryon number is not conserved. In the standard model the baryon
number is classically conserved but can be violated by quantum
mechanical effects known generically as ``chiral anomalies". The
process leading to matter creation is called ``spectral flow" and
can be pictured as a process in which particles flow from negative
energies to positive energies under the influence of an external
force. In this way real observable positive energy particles can
appear without simultaneous creation of antiparticles. Fig.~1(b)
illustrates a simple example of spectral flow occuring for
massless particles with electric charge $q$ (in units of
the charge of the electron) moving in a magnetic field. The
application of an electric field ${\bf E}$ leads to the
production of particles from the vacuum at the rate
\begin{equation}
\dot{n}=\partial_\mu j^\mu ={q^2 \over {4\pi^2}} {\bf E} \cdot {\bf
B} \ 
\label{1}
\end{equation}
per unit volume; factors of $(e/\hbar)$ have been absorbed in
the definition of the electric and magnetic fields.
This is an anomaly equation for the production of particles from
the vacuum of the type found by  Adler\cite{Adler1969} and by Bell and
Jackiw\cite{BellJackiw1969} in the context of neutral
pion decay. We see that for particle
creation it is necessary to have an asymmetric branch of the
dispersion relation $E(p)$ which crosses the axis from
negative to  positive energy.  We call such a branch a zero
mode branch; a spectrum of this type was first found for vortex core
excitations in a superconductor\cite{Caroli}.

Similar zero mode branches exist on a cosmic electroweak string
(also known as a Z-string), which is a structure of the Higgs
field that may have been produced during the  
electroweak phase transition.  The Higgs field gives the
particles mass outside the string core. This field vanishes
on the string axis and the fermions (quarks and leptons) living in the
core of the string behave like massless one-dimensional
particles. Spectral flow on a Z-string leading to production of
baryons from the vacuum with conservation of electric charge is
illustrated in Fig. 1(c).  Motion of the string across a
background electromagnetic field \cite{ewitten} or the
de-linking of two linked loops \cite{tvgf,jgtv}
provides a mechanism for cosmological baryogenesis \cite{barriola}
and could lead to the presence of antimatter in cosmic rays \cite{gstv}.

We should point out that baryon number violation is only one 
ingredient in a
cosmological baryogenesis scenario. The other ingredients are
that the system has to be out of thermal equilibrium, and charge
(C) and charge-parity conjugation (CP) symmetries should be
violated; these conditions are known as the ``Sakharov
conditions'' for cosmological baryogenesis \cite{sakharov}. 
In condensed matter the analogous symmetry
breaking is provided by rotation (for $^3$He) or a magnetic field
(for superconductors), and disequilibrium is provided by the 
motion of vortex lines.  We shall see that excitation momentum is the
analogue of baryon number.



The superfluidity of $^3$He is due to the formation of bound pairs of
$^3$He atoms known as Cooper pairs. There are two main superfluid
phases, A and B, which have different Cooper pair wave functions.
At the lowest temperature the superfluid vacuum state is obtained
in which all the atoms are Cooper paired. By analogy with the
Dirac picture of the vacuum (Fig. 1(a)) the Cooper-paired atoms are
seen as filling all the negative energy states and are
separated from the world of normal particles by the Cooper pair
binding energy. The Cooper pair wave function (usually called the
order parameter) which is responsible for this energy gap is thus
analogous to the Higgs field in electroweak theory since it is the
finite particle mass produced by the Higgs field that results
in the energy gap in electroweak theory between the negative and
positive energy states in Fig. 1(a). At higher temperatures in $^3$He 
some atoms are
excited from the ground state by the breaking of Cooper pairs and these
normal (non-superfluid) excitations interact with the
order parameter in a way that is similar to the interaction of the
standard model particles with the electroweak gauge and Higgs
fields. Thus the behaviour of leptons and quarks on fixed gauge
and Higgs field backgrounds can be modelled by the behaviour of
$^3$He excitations on fixed order parameter backgrounds such as
quantized vortices. In $^3$He a zero mode branch exists
for particles in the core of quantized vortices (Fig.~2(a)). 
A physically important 
charge in  $^3$He-A, $^3$He-B and superconductors which, like
baryonic charge in the standard model, is not conserved due to
the anomaly, is excitation momentum.  Spectral flow 
along the zero mode branch leads to an additional ``lift'' force
on a moving vortex (Fig.~2(b), the analogue of Fig.~1(c)).

This is most easily seen for the doubly
quantized continuous vortex (winding number $N=2$) in the A-phase
of superfluid $^3$He, which is the closest analogue of a 
Z-string.  The Cooper pairs in $^3$He-A have angular
momentum $\hbar$ and locally all the pairs have this angular
momentum aligned along a direction ${\hat {\bf l}}$.
The $N=2$ vortex is characterized
by a continuous distribution of the order parameter vector ${\hat{\bf l}}$ as
shown in Fig.~2(c). The time and space dependent ${\hat{\bf l}}$ vector
associated with the motion of the vortex produces a force on the excitations
equivalent to that of an ``electric field''
${\bf E}=k_F \partial_t {\hat {\bf l}}$ and a ``magnetic field''
${\bf B}=k_F{\bf \nabla}\times {\hat {\bf l}}$ acting on particles of
unit charge, where $k_F= p_F/\hbar$
and $p_F$ is the Fermi momentum.  Equation~(\ref{1}) can then be 
applied to calculate the rate at which left-handed
quasiparticles and right-handed quasiholes are created by spectral flow.
Since both types of
excitation have momentum $p_F{\hat {\bf l}}$, excitation momentum is
created at a rate
\begin{equation}
\partial_t {\bf P}= {1\over {2\pi^2}}\int d^3r~ p_F\hat {\bf l} ~(
{\bf E} \cdot   {\bf B} \, \, ) ~~.
\label{3}
\end{equation}
However, total linear momentum must be conserved.
Therefore Eq. (\ref{3}) means that,  in the presence of a
time-dependent texture, momentum is transferred from the
superfluid ground state (analogue of vacuum) to the heat
bath of excitations forming the normal component (analogue of matter).



Integration  of the anomalous momentum
transfer in Eq.(\ref{3}) over the cross-section
of the moving vortex gives the loss of linear momentum
and thus the additional force per unit length acting on the vortex  due to 
spectral flow:
\begin{equation}
{\bf F}_{\rm sf }= \partial_t {\bf P}=\pi \hbar N C_0
\hat{\bf z}\times ({\bf v}_n-{\bf v}_L) ~~.
\label{4}
\end{equation}
Here $\hat {\bf z}$ is the direction of the vortex,
$C_0= k_F^3/3\pi^2$,
${\bf v}_L$ is the velocity of the vortex line,
${\bf v}_n$ is the heat bath velocity, and
$N$ is the winding number of the vortex.

The insets to Fig.~3(a) show how the force on a vortex is
measured experimentally.  A uniform array of vortices is produced
by rotating the whole cryostat, and oscillatory superflow perpendicular
to the rotation axis is produced by a vibrating diaphragm, while 
the normal fluid (thermal excitations) is clamped by viscosity. The velocity
${\bf v}_L$ of the vortex array is determined by the overall
balance of forces acting on the vortices, conventionally written
as\cite{KopninVolovik1995,Bevan}
\begin{equation}
n_s \pi \hbar N \left[\hat{\bf z}\times ({\bf v}_L-{\bf v}_s)+
d_\perp \hat{\bf z}\times({\bf
v}_n-{\bf v}_L)+d_\parallel({\bf v}_n-{\bf v}_L)\right] =0~~,
\label{5}
\end{equation}
where $n_s(T)$ and ${\bf v}_s$ are the density and velocity
of the superfluid component. 
The first term is the Magnus force, which appears when the
vortex moves with respect to the superfluid and the terms
with $({\bf v}_n-{\bf v}_L)$ represent the nondissipative
transverse and  frictional longitudinal forces, proportional to dimensionless
parameters $d_\perp$ and $d_\parallel$ respectively, which appear if
the vortex moves with respect to the normal fluid.
The diaphragm has to provide a force equal and opposite to the Magnus
force to drive the superfluid. 
Measurement of the damping of the diaphragm resonance and of the 
coupling between the two orthogonal modes illustrated in Fig.~3(a)
enables both $d_\perp$ and $d_\parallel$ to be deduced.

For the
A-phase the spectral flow force in Eq.~(\ref{4}) combines with 
the Iordanskii force\cite{KopninVolovik1995} to give $d_\perp\approx
(C_0-n_n)/n_s$, where $n_n(T) = n -n_s(T)$ 
is the density of the normal component, with $n$ the total density. 
Since $C_0\approx n$, one has $d_\perp\approx 1$.  Our $^3$He-A
experiments made at 29.3bar and $T>0.82T_c$ are consistent with
this within experimental uncertainty: we find that
$|1-d_\perp|<0.005$\cite{Manninen}.

Any Fermi superfluid becomes similar to the A-phase 
in the vicinity of the vortex core.  However,
in $^3$He-B the spacing $\hbar\omega_0$ between bound states
on the anomalous branch in Fig.~2(a) becomes larger than the lifetime 
broadening $\hbar\tau^{-1}$ at low $T$ \cite{KopninVolovik1995}. 
There should thus be a transition from full spectral flow as
$T\rightarrow T_c$ to totally suppressed spectral flow as
$T\rightarrow 0$, a further aspect of the theory which can be tested
experimentally.  The interpolation
formula as a function of the relaxation parameter
$\omega_0\tau$ and the spectral flow parameter
$C_0\approx n$ \cite{KopninVolovik1995,Stone} can be written
in the form
\begin{equation}
d_\parallel-j(1-d_\perp)=
\frac{1}{\alpha+j(1-\alpha')}=
\frac{n}{n_s}\frac{\omega_0\tau}{1+j\omega_0
\tau}\tanh\frac{\Delta(T)}{2k_BT}~~,
\label{6}
\end{equation}
where $j=\sqrt{-1}$ and the $\alpha$ parameters are what are directly measured
experimentally\cite{Bevan}. The experimental results are compared 
with Eq.~(\ref{6}) in Fig.~3.   
The agreement is excellent in view of the approximations in 
the theory.


Our results thus show that the chiral anomaly is
relevant for the interaction of condensed matter vortices
(analogue of strings) with fermionic excitations (analogue
of quarks and leptons); this gives a firmer footing to chiral
anomaly calculations for baryogenesis on non-trivial backgrounds
of the Higgs field such as cosmic strings.  Superfluid $^3$He
is the most complex field theoretic system available in the
laboratory, and is therefore the closest experimental analogue of the
field theoretic foundations of the physics of the early universe.
It is characteristic of non-linear theories that experiment often
reveals qualitatively new phenomena that would be hard to find by
contemplation of the equations alone.  We may therefore hope that
imaginative experiments on superfluid $^3$He will generate new
ideas in cosmology.



ACKNOWLEDGEMENTS. The Manchester authors thank Mike Birse for advice  on field
theory and EPSRC for financial support. TV was partially supported
by the US Department of Energy.

CORRESPONDENCE to Henry.Hall@man.ac.uk
\pagebreak

FIGURE CAPTIONS                                      

Fig.~1: (a) Particles and antiparticles in the Dirac picture;
the thick line shows occupied negative energy states.  Promotion
of a particle from negative energy to positive energy creates
a particle-antiparticle pair from the vacuum.

(b) Spectrum of massless right-handed particles in a magnetic field
${\bf B}$ along $z$;
the thick lines show the occupied negative-energy states. 
The right-handed chirality of the particles means that their spin is
aligned with their linear momentum. Motion of the
particles in the plane perpendicular to ${\bf B}$ is
quantized into the  Landau levels
shown.  The free motion is thus effectively reduced to
one-dimensional motion along ${\bf B}$ with momentum $p_z$.
Because of the chirality of the particles  
the lowest ($n=0$) Landau
level, for which $E=cp_z$, is asymmetric: it crosses zero only 
in one direction. If
we now apply an electric field ${\bf E}$ along $z$, 
particles are pushed from negative to positive energy
levels according to the equation of motion $\dot p_z=qE$,
and the whole Dirac sea moves up, 
creating particles and electric charge from the vacuum.
This motion of the particles along the
``anomalous'' branch of the spectrum is called
spectral flow. The rate of particle production is
proportional to  the density of states at the Landau level,
which is  $\propto q\vert {\bf B}\vert$, so that the rate of
production of particles from the vacuum is
$\propto q^2 {\bf E} \cdot {\bf B}$.

(c)  Anomalous branches of the spectrum of u-quarks ($q=+2/3$)
and d-quarks ($q=-1/3$) in the core of a Z-string. 
There are anomalous branches for electrons ($q=-1$) and neutrinos 
($q=0$) also. The neutrino and u-quark propagate in one
direction along the string while the
electron and d-quark propagate in the opposite direction. 
For every electron produced, two u-quarks
and one d-quark are created so that there is no net production of
electric charge but the baryon number $B$ increases by one (2 u-quarks
+ 1 d-quark $\equiv$ 1 proton = 1 baryon). 

Fig.~2:  (a) Spectrum of particles in the core
of vortices in $^3$He and superconductors;  filled circles 
show occupied states.  The only
difference from the Z-string is that the anomalous
branch, $E_0(p_z,L_z)=-L_z\hbar\omega_0(p_z)$, of the spectrum
crosses zero as a function of the discrete angular momentum
$\hbar L_z$, where $L_z$ is a  half-odd integer.  Typically
the interlevel distance $\hbar\omega_0$ is
very small compared to the characteristic energy scales in
superconductors and Fermi superfluids, and can be comparable
to the level width $\hbar/\tau$ resulting
from the scattering of core excitations by free excitations in
the heat bath outside the core. If $\omega_0\tau < 1$ the
levels overlap and spectral flow is allowed. This type of 
spectral flow is analogous to ``hopping conduction'' in a solid; 
it is assisted rather than impeded by collisions.

(b) Such spectral flow is induced by the motion of the vortex with   
respect to the heat bath.
If the vortex moves with velocity $v_x$ along $x$ the rate of change
of excitation angular momentum in the moving core is ${\dot L_z}=
{\dot x}p_y=v_x p_y$.  The levels cross zero at a rate
$-{\dot L_z}/\hbar$ leading to spectral flow in opposite senses
for $p_y>0$ and $p_y<0$ as indicated.
Since the left-handed quasiparticles created for $p_y>0$ are equal
in number to the right-handed quasiholes created for $p_y<0$
there is no net production of chiral charge $C$, but excitation
momentum is created at a rate ${\dot p_y}=-v_x{p_y}^2/\hbar$, 
independent of the sign of $p_y$ (shown for negative $v_x$).

(c) An example of an $N=2$ continuous vortex in $^3$He-A. The cones  
indicate the local direction of the order parameter vector ${\hat{\bf l}}$,
which is parallel to the angular momentum of
the Cooper pairs, and is analogous to the direction of the isotopic spin
in electroweak theory. Rotation of the cones about ${\hat{\bf l}}$
indicates a change in phase of the order parameter; note the $4\pi$ phase
change around the perimeter of the diagram, corresponding to two quanta
of anticlockwise circulation.  Four topologically different
types of $^3$He-A vortices have been observed \cite{Parts}.

Fig.~3: (a){\it Upper inset} Experimental cell. The aluminized Kapton film
diaphragm separates two disc-shaped regions of superfluid $^3$He,
each 100$\mu$m thick.
The roof of the cell has six electrodes set into it by
means of which the oscillations of the diaphragm may be driven and
detected electrostatically. In the oscillating modes of interest the
motion of the diaphragm displaces the superfluid as indicated, while
the normal component of the fluid (the heat bath) is clamped by its
high viscosity. Rotation at angular velocity $\Omega$  about a vertical
axis produces vortices  normal to the diaphragm.  These vortices
produce additional dissipation proportional to $\alpha\Omega$ and
coupling between two orthogonal modes of the diaphragm, with displacement 
patterns shown in the {\it lower inset}, proportional to
$(1-\alpha')\Omega$. The measured spectral flow parameters
are related to the parameters in
Ref.~\cite{Bevan} by $\alpha=B\rho_n/2\rho$ and
$\alpha'=B'\rho_n/2\rho$.

{\it Main frame}
The measured $(1-\alpha')/\alpha$ for $^3$He-B at 20bar is
compared with the theoretical value $(1-\alpha')/\alpha\approx
\omega_0\tau$. The temperature dependence of  $\omega_0\tau$ is not
very well known, since it is sensitive to the details of the
quasiparticle scattering. It can be estimated by the formula $\omega_0\tau
=(a\Delta(T)/k_BT)^2\exp (\Delta(T)/k_BT)$ \cite{KopninVolovik1995}.
The solid line is for $a=0.067$.  The fit is much improved if an effective 
energy gap 0.6 of the bulk value is assumed; the dashed line is for
$\Delta(T)=0.6\Delta_{\rm bulk}(T)$ and $a=0.214$.  But the justification
for such an assumption is not clear.

(b)  The experimental parameter $(1-\alpha')$ for $^3$He-B at 20bar is
compared with the `spectral flow' prediction of
Eq.~(\ref{6}): $(1-\alpha')\approx
n_s(T)/(n\tanh(\Delta(T)/ 2k_BT))$ (solid line). Note that this fit
is independent of the relaxation parameter $\omega_0\tau$.  The
dashed line is for $\Delta(T)=0.6\Delta_{\rm bulk}(T)$, and fits much better
below $0.6T_c$.  This fit shows that spectral flow is less strongly
suppressed at low temperatures than the full energy gap would suggest.


\pagebreak
\includegraphics[width=150mm]{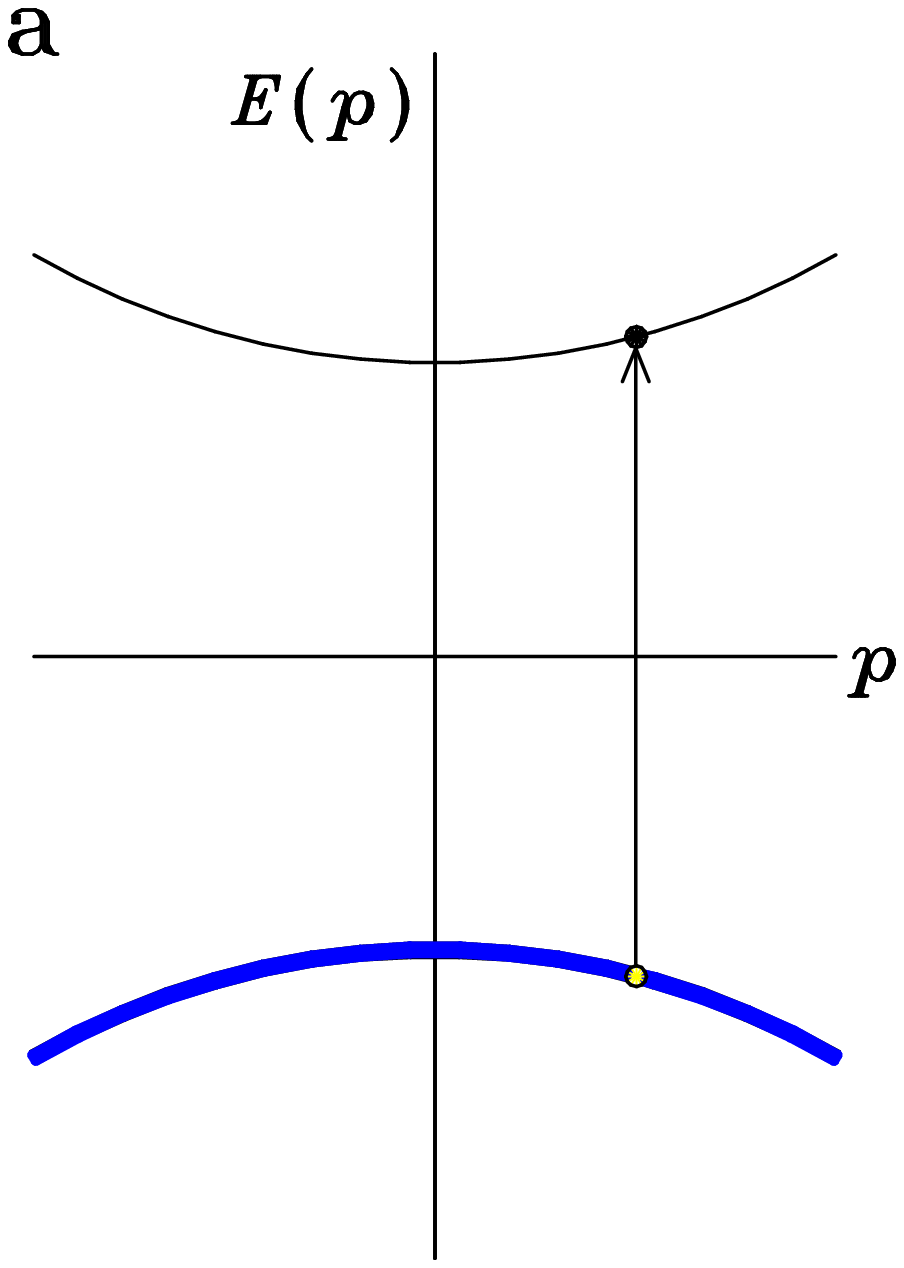}\\ 
Figure~1(a)\\

\pagebreak
\includegraphics[width=150mm]{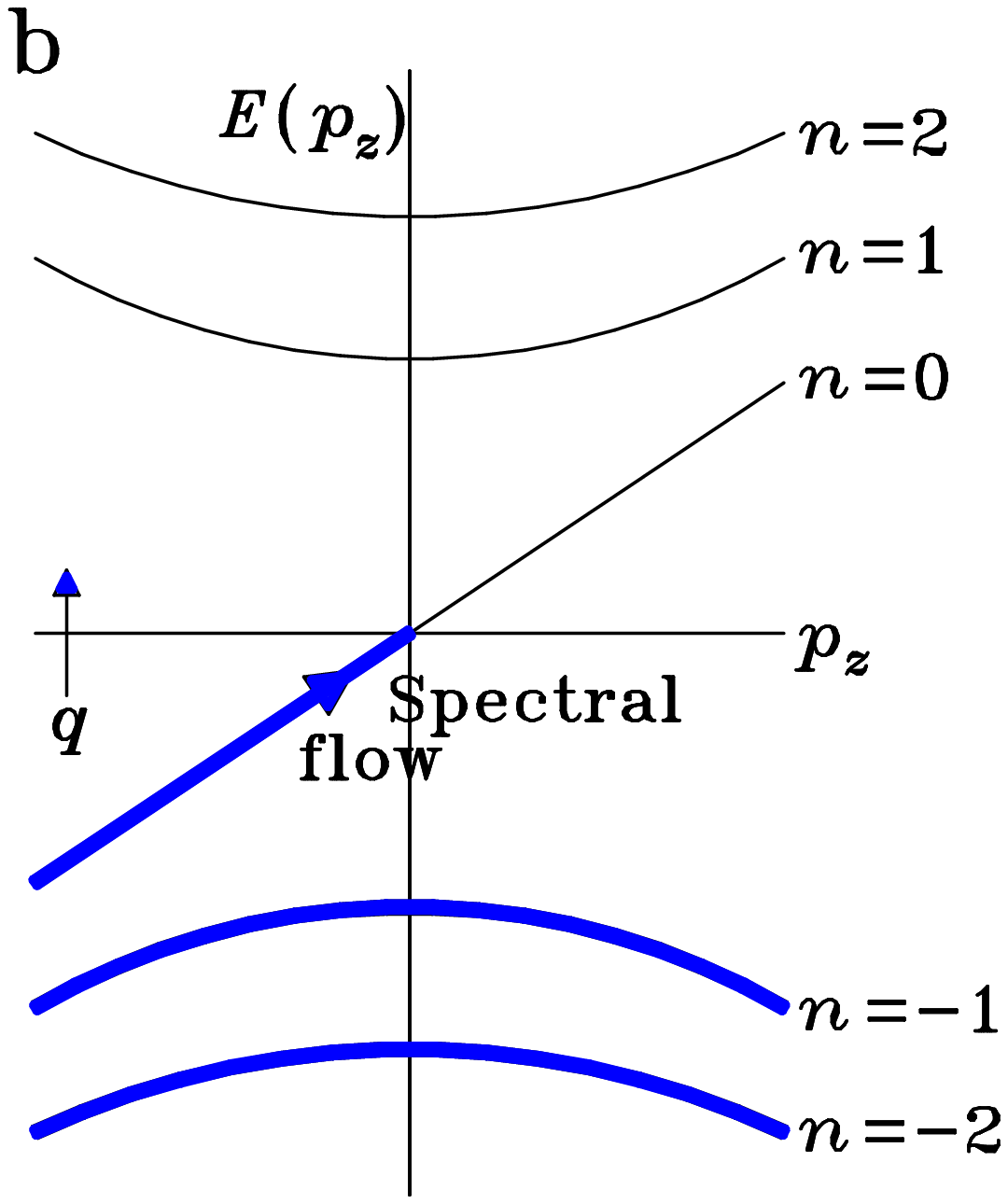}\\ 
Figure~1(b)\\

\pagebreak
\includegraphics[width=150mm]{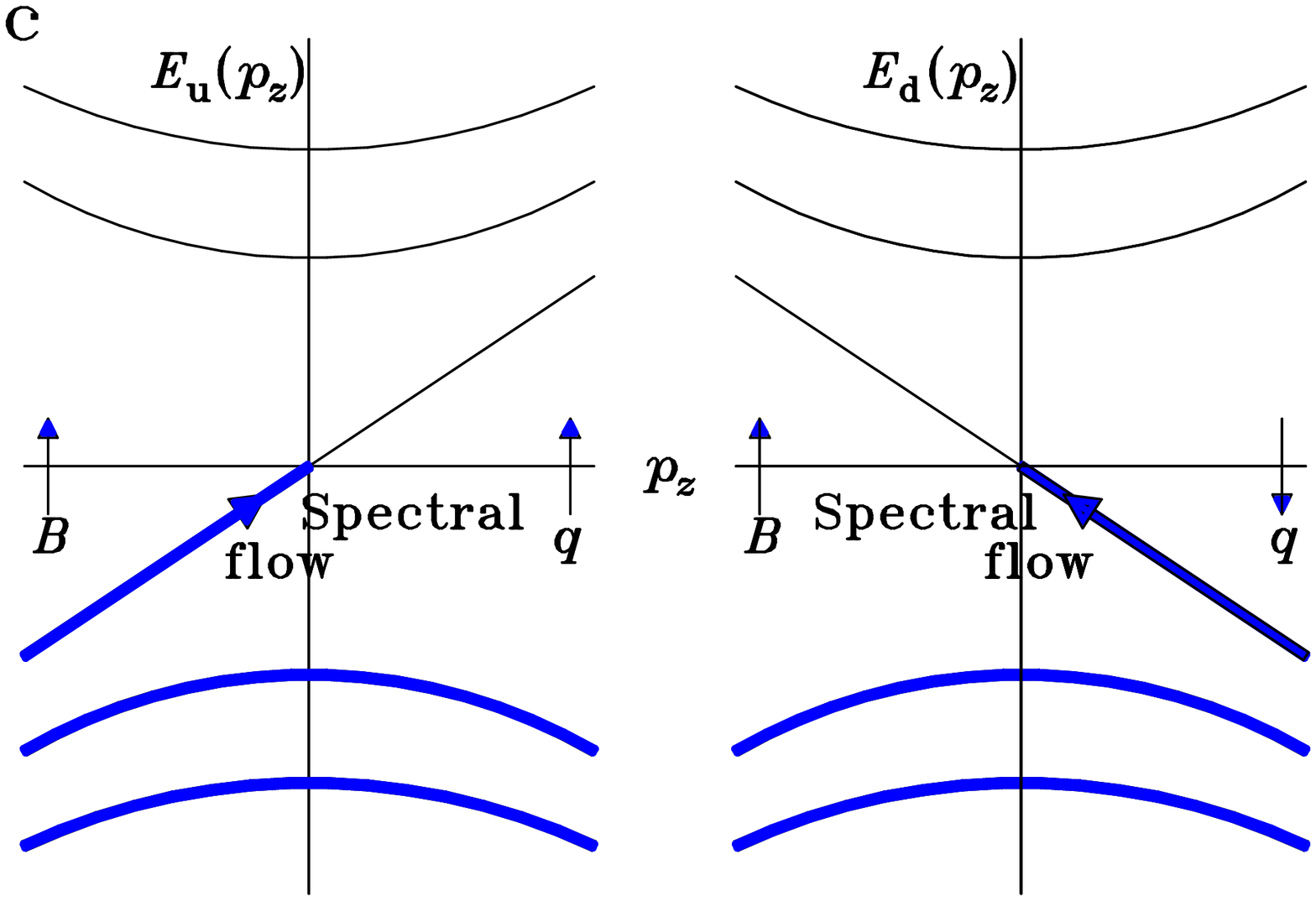}\\ 
Figure~1(c)\\

\pagebreak
\includegraphics[width=150mm]{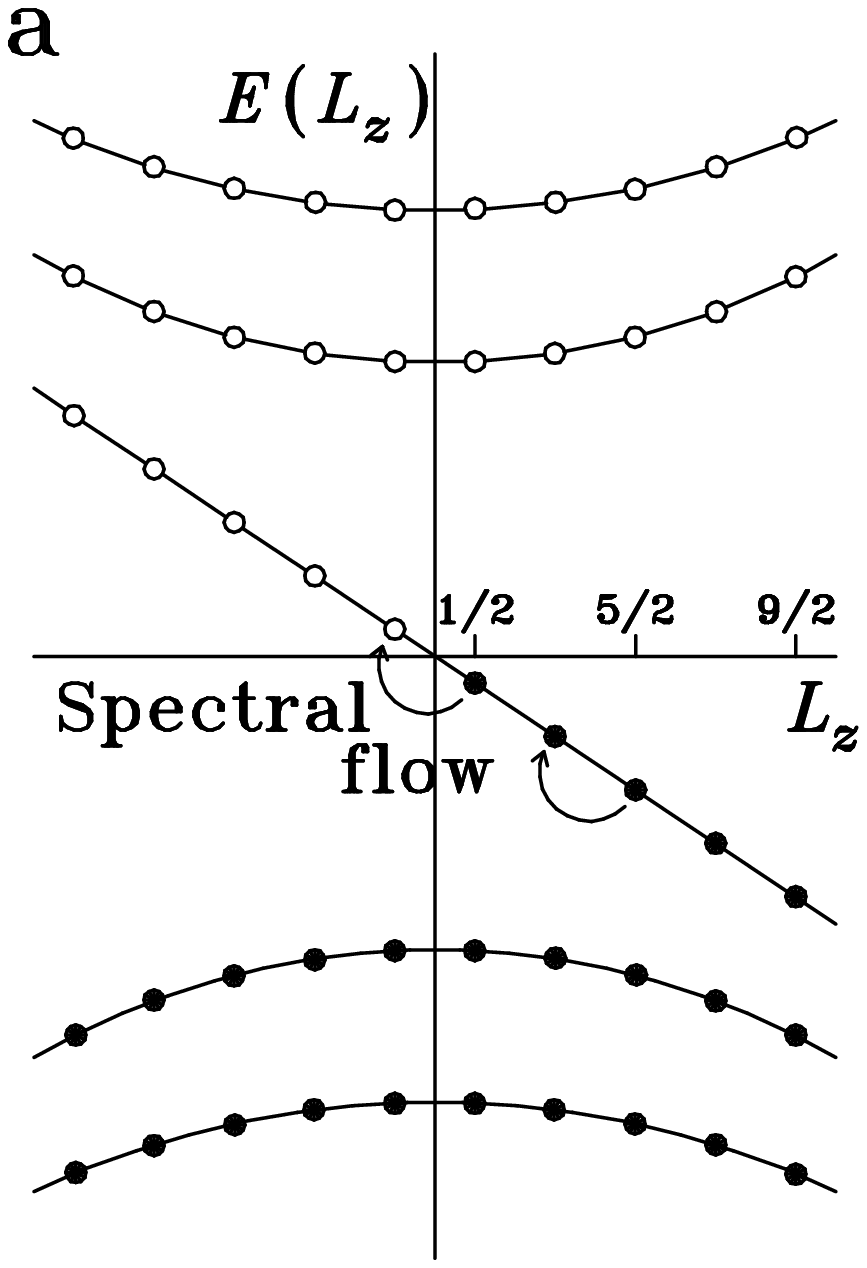}\\ 
Figure~2(a)\\

\pagebreak
\includegraphics[width=150mm]{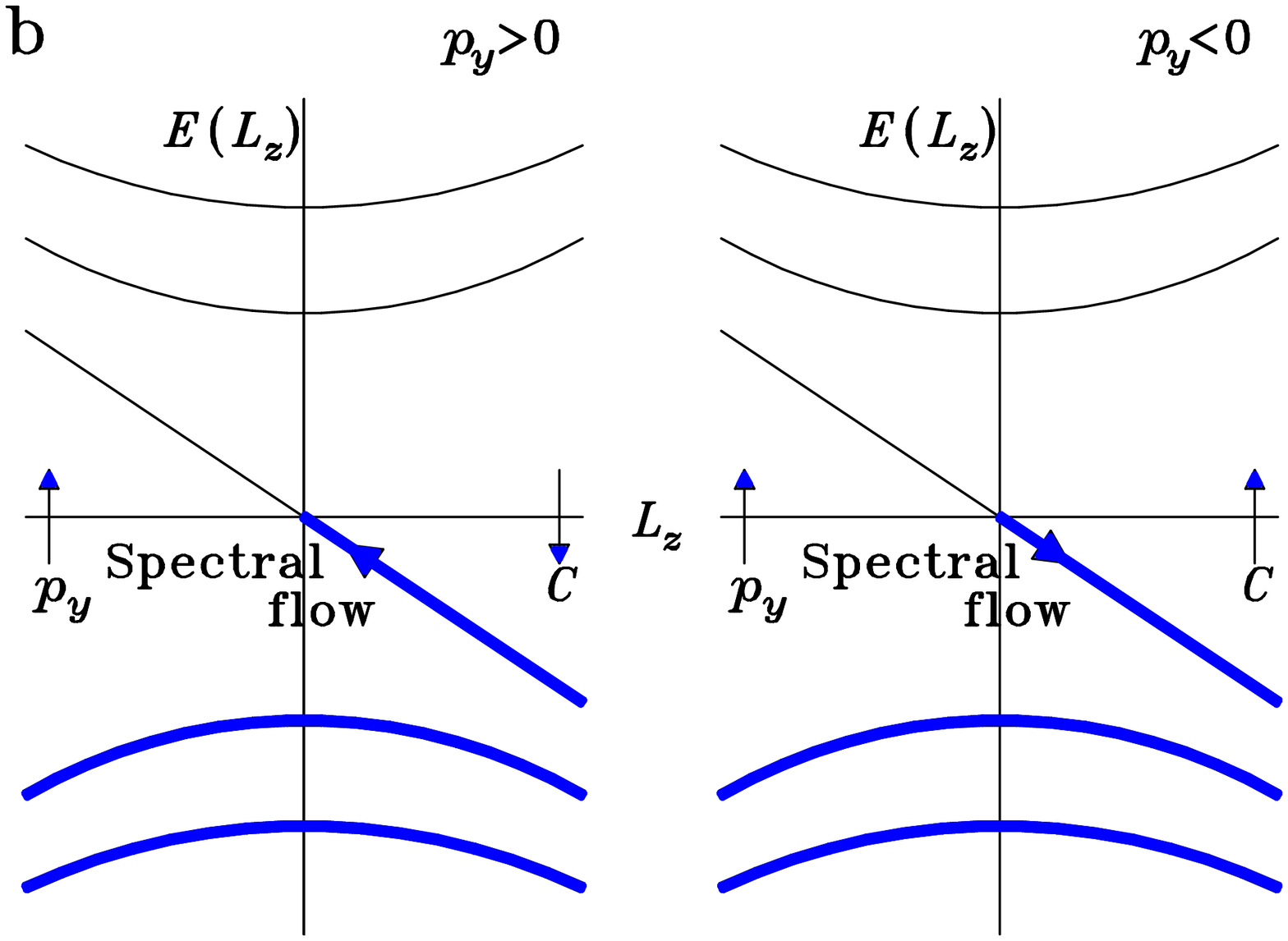}\\ 
Figure~2(b)\\

\pagebreak
\includegraphics[width=150mm]{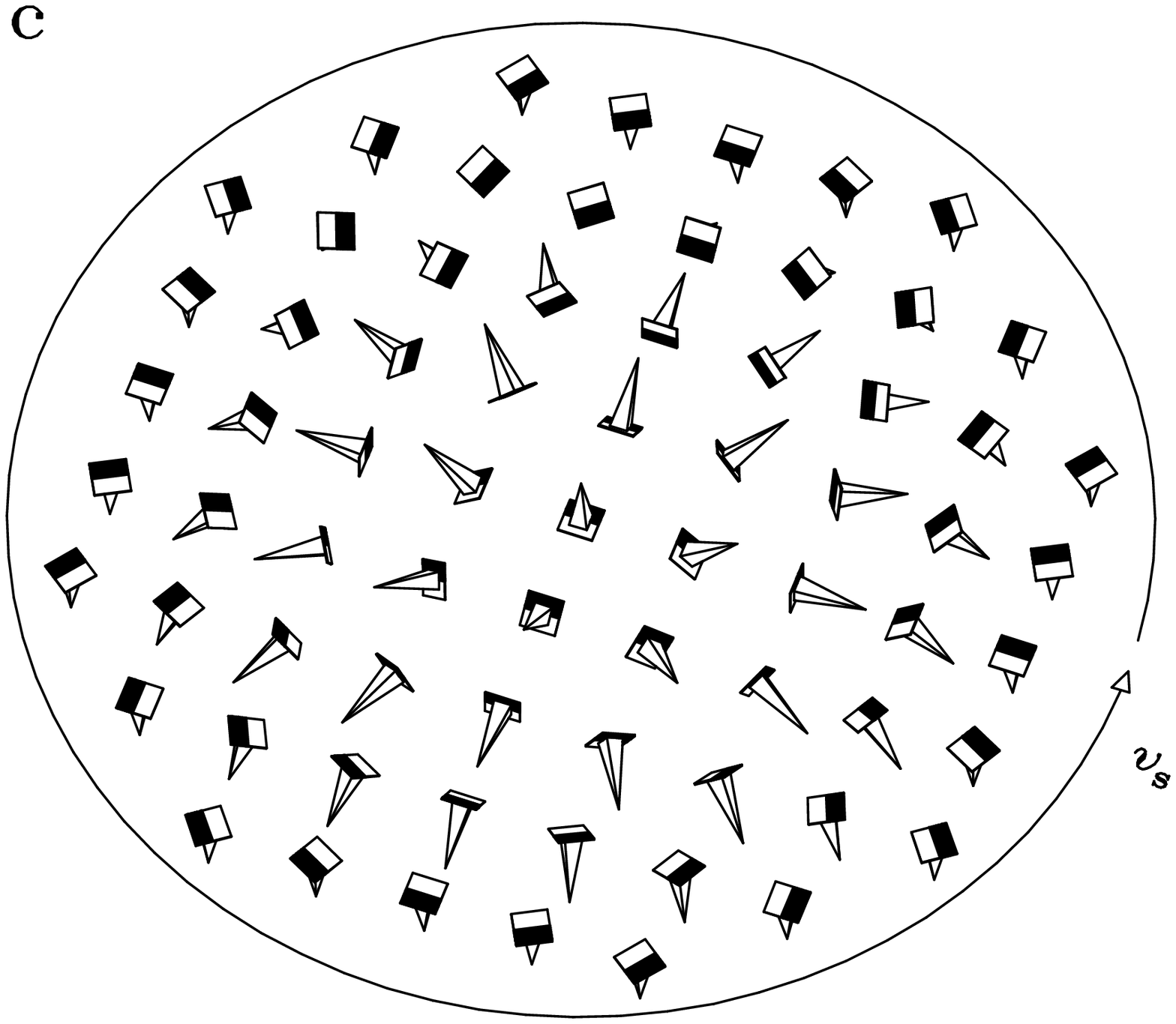}\\ 
Figure~2(c)\\

\pagebreak
\includegraphics[width=150mm]{natfig3a.ps}\\ 
Figure~3(a)\\

\pagebreak
\includegraphics[width=150mm]{natfig3b.ps}\\ 
Figure~3(b)\\

\end{document}